# Why does superfluid helium leak out of an open container?

V. A. Golovko

Moscow State Evening Metallurgical Institute

Lefortovsky Val 26, Moscow 111250, Russia

E-mail: mgvmi-mail@mtu-net.ru

Abstract

The fact that superfluid helium always leaks out of an open container is usually explained by the phenomenon of wetting. In the present paper it is demonstrated that this explanation is unconvincing. The fact can be readily explained from the viewpoint of the interpretation of superfluidity proposed earlier by the author according to which superfluidity is an equilibrium state of liquid helium where the symmetry is spontaneously broken because of an intrinsic superflow. Experiments on the thickness of moving helium films that have given rise to much controversy are discussed as well. Some experiments concerning the phenomena considered in the paper are proposed.

One of the unusual properties of superfluid helium-4 (helium II) is the fact that, when placed in a container open at the top, it always crawls up the wall and out of the container [1]. This fact is usually ascribed to the phenomenon of wetting owing to which the whole of the surface of the container is coated with a helium film (this is clearly seen from, e.g., [2]). First of all, we shall demonstrate that wetting cannot explain the fact.

Let us discuss the process of wetting. At the outset when the container is just filled with helium, a force **F** arising from interaction between helium atoms and molecules of the container's wall (for an overview of relevant forces see, e.g., [3]) draws the helium film upwards as shown in Fig. 1. It should be emphasized that the force **F** parallel to the wall acts solely on the edge of the film since in the interior of the film far away from any boundaries no force parallel to the wall can exist by symmetry considerations. It is to be added that the symmetry considerations concern, of course, films of uniform thickness alone. If the wall is vertical as in the present case, the equilibrium film thickness decreases with height [1]. The resulting force compensates the weight of the vertical film. Intermolecular forces that are perpendicular to the wall and attract the film to the wall are not represented in the figure.

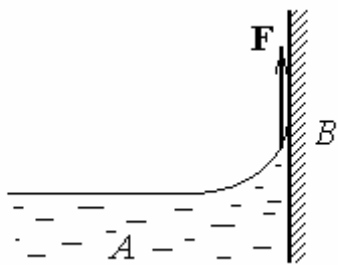
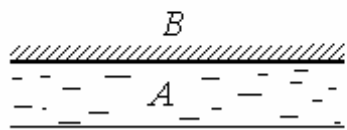

**Fig. 1.** Outset of wetting.   **Fig. 2.** End of wetting.
*A*: helium, *B*: wall of the container.   *A*: helium film, *B*: bottom of the container.

When the film covers the entire surface of the container (the outer surface inclusive) and all edges of the film disappear as in Fig. 2, the driving force **F** vanishes. The leakage of helium out of the container should stop at this stage inasmuch as there are no longer forces that would draw helium upwards overcoming the force of gravity (the absence of viscosity plays no role here). As a result, the amount of helium that escapes from the container in this situation will be small and equal to the one in the film. The experiment shows the opposite: helium II continues to creep up the wall against the force of gravity forming drops that fall from the container's bottom until the container is empty [1,2].

Before proceeding further let us make a general remark. When discussing helium films Dzyaloshinskii *et al.* [3] remark that temperature corrections to the equations obtained should be relatively small so that the profile of the helium film should be essentially the same below and above the λ-point (outside the immediate vicinity of the λ-point). Viscosity cannot stop a flow resulting from a force acting in a liquid, it can only slow down the flow. Consequently, if there is a force that could overpower gravity and the force is due to wetting, the phenomenon of the open container under discussion must be observed above the λ-point as well where helium is not superfluid. Moreover, the phenomenon might be observed with some classical liquids too as long as no quantum effects figure in the reasoning concerning Figs. 1 and 2. However only with helium and only below the λ-point does one observe the phenomenon. Therefore, the phenomenon is closely connected with superfluidity.

Let us discuss nevertheless possible causes for the helium leakage out of the container without account taken of superfluidity. The lower part of the helium film shown partially in Fig. 2 has a weight that exerts a pressure on the lower surface of the film. However the gravitational force that acts on helium atoms is of secondary importance with respect to ordinary intermolecular forces that determine the thickness of the film and is always discarded when considering horizontal films. Although the pressure due to the weight of the helium film should be negligible, it can be readily calculated to be $p_w = \rho g \Delta h$ where $\rho$ is the liquid density, $g$ is the gravitational acceleration and $\Delta h$ is the thickness of the film in Fig. 2. We shall return to this result later on.

Besides the pressure due to the weight of the film, the pressure in the lower part of the helium film is higher than in the container owing to the hydrostatic pressure $p_h = \rho g h$ where $h$ is the height of the helium level in the container with reference to the lower part of the film (Fig. 2) where $p_h$ is a maximum. Helium will leak out of the container only if the hydrostatic pressure tears the helium film. The last will occur if the pressure $p_h$ is capable of breaking the balance of the intermolecular forces that determine the film thickness. We can get a rough idea of the magnitude of these forces in liquid helium in the following way. The attractive intermolecular forces create an internal pressure $p_i$ added to the external pressure $p$. If we take, as an example, the van der Waals equation of state, then $p_i = a/V_m^2$ where $a$ is the van der Waals constant and $V_m$ is the volume of one mole. The constant $a$ can be computed with use made of the critical temperature and pressure in the usual fashion, which gives $a = 3.46 \cdot 10^{-3}$ Pa·m$^6$/mol$^2$ for helium (a similar value can be found in handbooks). With the helium II density $\rho = 0.145$ g/cm$^3$ the molar volume is $V_m = 27.6$ cm$^3$/mol. As a result, for the internal pressure in helium II we obtain $p_i = 44.8$ atm, whereas one atmosphere corresponds to the height of the column of helium equal to $h = 71.2$ m with the above density. Thus we see that, for actual values of $h$, the hydrostatic pressure is very small as compared to the intermolecular forces that determine the film thickness, and thereby this pressure can in no way tear the film. We see also that the pressure $p_w = \rho g \Delta h$ due to the weight of the lower part of the helium film discussed above is negligible even when compared with the hydrostatic pressure $p_h = \rho g h$ seeing that $\Delta h \ll h$.

We turn now to the question as to how superfluidity affects the behaviour of helium in the container. If helium II leaks out of a container through a narrow capillary (through the film in our case), a rise in temperature occurs in the container [4]. On the other hand, the superfluid fraction (while only it can move in the film) tends to move toward the region where the temperature is higher [1], into the container in the present instance. Therefore, the above heating of helium II in the container should hinder its leakage out of the container. It should be stressed the difference between the isolated container under consideration and a container placed in a reservoir filled with helium II (see Fig. 3a below). In the latter case, heat readily escapes from the container through its walls because the thermal conductivity of helium II in the reservoir is very high [1]. Cooling is especially efficacious when the container is metallic. The isolated container is surrounded by a vapour whose thermal conductivity is poor, so that heat remains in the container and should block the leakage of helium out of the container according to the above. Therefore superfluidity only impedes the helium leakage in addition to gravity.

Yet another inexplicable fact exists. Even if, for unknown reasons, the superfluid fraction acquires a velocity owing to which it is able to quit the container through the film, a normal (nonsuperfluid) fraction is also present in the container, the fraction that cannot move in the film and moreover which does not at all move. The question arises as to why all helium escapes from the container, the normal fraction inclusive. As to the superfluid fraction it may be added that, if the fraction with the above velocity quits the container, the fraction will pass under the container's bottom through the film shown in Fig. 2 and will return into the container from the opposite side without any leakage.

Thus we see that the phenomenon of wetting and the existing view on superfluidity cannot explain why superfluid helium leaks out of the open container.

In Ref. [5] (see also [6,7]) a new interpretation of superfluidity different from the well-known Landau mechanism was proposed. Superfluidity is interpreted as an equilibrium state of liquid helium in which the symmetry is spontaneously broken because of an intrinsic superflow whose magnitude is fixed by thermodynamics. Symmetry breaking occurs at the λ-transition analogously with a ferromagnet whose symmetry is broken at the Curie point because of the appearance of spontaneous magnetization.

The existence of spontaneous superflows readily explains the enigma of the open container. It is clear that the superflow tries and chooses a rectilinear trajectory and for this reason moves upwards along the wall in the film and eventually out of the container despite the force of gravity and the adverse temperature effect discussed above. It is essential that the superflow moves upwards across the whole perimeter of the internal wall, so that the superflow cannot return into the container from the opposite side accumulating inevitably at the lower surface of the container's bottom. Once enough of helium leaks out of the container, at this surface there will be a bulk liquid rather than the film discussed above. The bulk liquid cannot be located at the lower surface of the bottom. It forms drops that will fall from the container's bottom while helium will continue to quit the container until the last is empty. It is to be emphasized that, although the superfluid fraction alone quits the container, the fraction cannot disappear from it leaving behind solely the normal fraction as long as the thermodynamic equilibrium requires the presence of the superfluid fraction in the container at the existing temperature. Therefore a new superfluid fraction is created in the container at the expense of the normal one and, what is important, the fraction created possesses a nonzero velocity characteristic of the thermodynamic equilibrium and thereby is capable of quitting the container whereas the amount of the normal fraction in the container diminishes. Owing to these causes, all helium escapes from the container, which is seemingly inexplicable at first sight as was mentioned above.

The following should be remarked as well. As discussed above there occurs heating in the container when the superfluid fraction escapes from it. Heat, however, passes through the container's bottom and goes away with the drops. It is interesting to note that when growing the drops are cooled as the superfluid fraction enters them, owing to an effect inverse to the one observed in the container [4]. The cooled drops pump heat out the container more efficaciously. Besides, some cooling is possible also because of evaporation of liquid helium.

In this context it is worthwhile to discuss experiments on the thickness of the helium film that have given rise to much controversy. Kontorovich [8] predicted that the moving film should be thinner than that at rest (see also [9]). However measurements by Keller [10] demonstrated that the thickness of the helium film was independent of its state of motion. Keller's results can be readily explained if the superflows are taken into account.

First of all, let us consider the observed flow of helium II between the container and a reservoir the container is placed in for different levels of helium in the container and reservoir (Fig. 3, see also [1]). Owing to the helium film the container and reservoir constitute communicating vessels while the superfluid fraction alone is able to flow in the film. If the levels are identical, there is no visible flow (Fig. 3c). In essence, Keller's experiment [10] consisted in measuring the thickness of the helium film in situations corresponding to Figs. 3a, 3b and 3c. He found that the thickness was the same in the static case where there was no visible flow (Fig. 3c) as well as in the case of the moving film (Figs. 3a and 3b). Furthermore, the thickness was independent of the flow rate determined from the speed of variation of the helium level in the container *C*. Keller's results appeared to be in clear disagreement with Kontorovich's calculations.

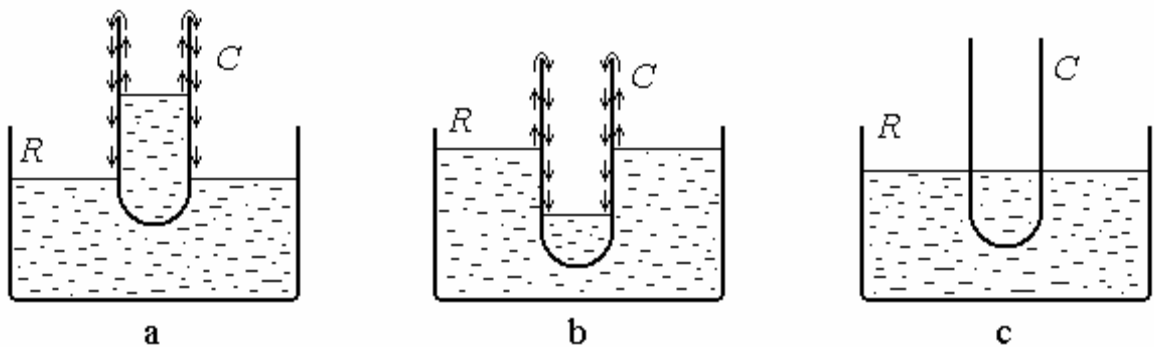

**Fig. 3.** Apparent flow of helium II over the wall of the container. *C*: container, *R*: reservoir.

It can be shown, however, that there is no disagreement if the spontaneous superflows are taken into account. As mentioned above the superflow always tries and moves along the wall. Besides, the equilibrium state of helium II in the film implies the existence of a spontaneous superflow as in the bulk. The velocity $v_s$ of the superflow is fixed by thermodynamics [5] and does not depend on the environment. We first turn to Fig. 3c. On each side of the container's wall, inside and outside, there are ascending and descending superflows forming a pattern similar to the one discussed in the concluding section of [5] for the case of a confined volume. Figuratively speaking, different parts of the film creep in opposite directions. The resultant flow is zero, so that helium is seemingly immobile. In the situations presented in Figs. 3a and 3b there are the ascending and descending superflows on both the sides of the wall as well. In the case of Fig. 3a, however, the resultant flow is directed from the container *C* to the reservoir *R* because the easiest way to reach the rim of the container is when the superflow goes from the container. Besides, the direction of the resultant flow is dictated by the fact that the container and reservoir are communicating vessels. Fig. 3b represents the inverse situation. In all the cases the velocity of the superflows in the film is equal to $v_s$ so that the film thickness must be one and the same as was observed by Keller [10]. Thus, in actual fact Keller's experiments support the existence of spontaneous superflows.

The apparent discrepancy between Keller's results and Kontorovich's prediction stimulated a large body of experiments on the thickness of helium films a review of which can be found in [11]. Some of the experiments confirmed Keller's results while the others were contradictory to the results demonstrating that the film thickness depended on the state of visible motion in agreement with the Kontorovich prediction although sometimes the agreement was qualitative rather than quantitative. The disparities in the experimental data reported can be explained as follows. The distribution of the spontaneous superflows in an experimental apparatus depends upon the geometry of the apparatus, the design of the experiment, experimental conditions. Even a sight tube with whose help one determines the level of helium may redistribute the superflows. It is not known how the pattern formed by the superflows behaves in different apparatuses under different experimental conditions, especially where time-dependent and other nonequilibrium processes are involved. How does application of heat used in some experiments affect the pattern? Besides, it is not clear whether, in one experiment or another, one measures the real velocity $v_s$ of the superflow or an averaged value. For example, as pointed out above when discussing Fig. 3c the averaged velocity is nil in this instance while $v_s \neq 0$. All these unknown factors may explain the disparities in the experimental data mentioned above.

Let us examine, as an example, the results of Kwoh and Goodstein [12] that are in seeming contradiction with Keller's results. The scheme of their first experiment is described by Fig. 3b but their reservoir *R* was small. They introduced sufficiently little helium into the reservoir that all of it emptied into the container *C*. As a result, the movement of helium along the wall of the container was similar to a free fall, so that to the velocity of the superflow $v_s$ was added the fall velocity. In accord with Kontorovich's reasoning the helium film should be thinner at that instant with respect to the thickness at the end of the experiment where helium was at apparent rest (in reality its local velocity was $v_s$), which was observed by Kwoh and Goodstein. In Keller's experiments, the processes evolved slowly and the superflow velocity was always equal to $v_s$. In their second experiment, Kwoh and Goodstein [12] started with enough helium in the reservoir and observed *U*-tube oscillations between the inside and outside levels in the container *C*. We cannot discuss this experiment because the behaviour of the pattern formed by the superflows in the course of such oscillations remains unknown. We restrict ourselves to quoting Kwoh and Goodstein [12] as to their second experiment: "Quantitative agreement with the theory is only approximate, for unknown reasons". It should be added that, when reading the discussion of other experiments on the thickness of helium films in Ref. [11], one can also meet words of the type "the experiment remains unexplained".

A remark needs to be made. As was mentioned above Dzyaloshinskii *et al*. [3] pointed out that the profile of the helium film should be essentially the same below and above the λ-point (outside the immediate vicinity of the λ-point). In actual fact the helium film is noticeably thinner below the λ-point [13,14]. This also can be explained by the existence of spontaneous superflows below the λ-point since even in a state of thermodynamic equilibrium the superflows move at a speed of $v_s$. At the same time, Kontorovich's calculations [8] should be refined in this case because the superflow not only adds a kinetic energy but also changes the pair correlation function [5] whereas this last function enters into formulae for various thermodynamic quantities. In the immediate vicinity of the λ-point, other factors play a role as well [13,14].

Closing the paper experiments concerning the phenomena considered in the present paper may be proposed. First of all, it is worthwhile to investigate how the rate of the helium leakage out of the container depends upon the temperature. The leakage rate should be proportional to the momentum **P** carried by the superflow. According to [5], **P** = $\rho_c \mathbf{p}_0 V$ where $\rho_c$ is the condensate density, $\mathbf{p}_0$ is a vector whose magnitude $p_0$ is determined by the thermodynamics of superfluid helium and $V$ is the volume of helium. At the same time it should be emphasized that only a part of superfluid fraction penetrates into the film and quits the container while the second part will circulate inside the container as discussed in [5]. The relation between these two parts may depend on the geometry of the container, the quantity of helium in the container and even on the magnitude of **P** at a given quantity of helium. At a given volume $V$ of helium in the container and under the assumption that the part of superfluid helium which penetrates into the film is fixed, the leakage rate will be dependent on $\rho_c p_0$ alone. Within the framework of the Landau phase transition theory applied to the present situation in section 3 of [7], the temperature dependence of $p_0$ can be neglected, in which case the leakage rate will increase proportionally to $\rho_c$ as the temperature lowers. If $p_0$ depends on the temperature perceptibly, the temperature dependence of the leakage rate can be more complicated. The behaviour of the leakage rate at a given temperature when the level of helium in the container lowers can be intriguing too seeing that in this process not only the volume $V$ decreases but there may also occur redistribution between the above two parts of the superfluid fraction.

It is of interest also to find out the maximum height up to which the spontaneous superflow is able to climb at a given temperature (the author is not aware of any experiment of this kind). The experiment can be simple. One should take a long beaker and fill it with helium II. Helium will begin to leak out of the beaker and its level will commence to lower. It is sufficient to measure the height from the level to the rim of the beaker when the lowering of the level ceases. It should be stressed that the spontaneous superflows will exist even when helium stops quitting the beaker. In the last instance the superflows move up and thereafter should turn down. It is interesting to note that the helium film just above the turning-points should be thicker than just below in line with Kontorovich's argument (see however the remark above as to Kontorovich's calculations). The thickening of the film will however occur only if there are no superflows above the turning-points. On the other hand, the spontaneous superflows can exist at the thermodynamic equilibrium of helium II even in the film. In this connection the pattern of spontaneous superflows in the upper part of the beaker may have peculiarities; in particular, local closed superflows may form there. The spontaneous superflows can disappear when, with increasing height, the film becomes so thin that conditions inside it will differ substantially from the conditions in the bulk of the liquid.